\documentstyle[sprocl,psfig]{article}
\def\Journal#1#2#3#4{{#1} {\bf #2}, #3 (#4)}

\def\be{\begin{equation}}
\def\ee{\end{equation}}
\def\bea{\begin{eqnarray}}
\def\eea{\end{eqnarray}}

\begin{document}
\title{{\em In: M.~Suzuki and N. Kawashima (eds.) ``Coherent Approaches to Fluctuations (Proc. Hayashibara Forum 95)'', pp. 59-64, World Scientific (singapore, 1995).}\vspace{1.5cm}\\ 
DYNAMIC ENTROPIES, LONG--RANGE CORRELATIONS AND FLUCTUATIONS IN
  COMPLEX LINEAR STRUCTURES}

\author{WERNER EBELING, ALEXANDER NEIMAN and THORSTEN P\"OSCHEL}

\address{Humboldt--Universit\"at zu Berlin, Institut f\"ur Physik,\\ 
  Invalidenstr. 110, D--10115 Berlin, Germany} 

\maketitle

\abstracts{
  We investigate symbolic sequences and in particular information
  carriers as e.g. books and DNA--strings. First the higher order
  Shannon entropies are calculated, a characteristic root law is
  detected. Then the algorithmic entropy is estimated by using
  Lempel--Ziv compression algorithms. In the third section the
  correlation function for distant letters, the low frequency Fourier
  spectrum and the characteristic scaling exponents are calculated. We
  show that all these measures are able to detect long--range
  correlations. However, as demonstrated by shuffling experiments,
  different measures operate on different length scales. The longest
  correlations found in our analysis comprise a few hundreds or
  thousands of letters and may be understood as long--wave
  fluctuations of the composition.
}
\section{Introduction}
The purpose of this paper is to compare different correlation measures
based on methods of statistical physics. We aim to analyse correlations 
and fluctuations comprising at least several hundreds of letters. 
The characteristic quantities which measure long correlations are dynamic
entropies~\cite{ebelingnicolis92,ebelingpoeschel,epa}, correlations
functions and mean square deviations, $1/f^\delta$
noise~\cite{voss,anishchenko}, scaling exponents~\cite{peng,stan},
higher order cumulants~\cite{ebelingneiman} and, mutual
information~\cite{likaneko,li}. Our working hypothesis which we
formulated in earlier papers~\cite{ebelingnicolis92,anishchenko}, is
that texts and DNA show some structural analogies to strings generated
by nonlinear processes at bifurcation points. This is demonstrated
here first by the analysis of the behaviour of the higher order
entropies. Further analysis is based on the mapping of the text to
random walk models as well as on the spectral analysis. The random
walk method which was proposed by Peng et al.~\cite{peng}, found
several applications to DNA sequences~\cite{pengbuldyrev,likaneko} and
to human writings~\cite{ebelingneiman,schenkel}.  In order to find out
the origin of the long--range correlations we studied the effects of
shuffling of long texts on different levels (letters, words,
sentences, pages, chapters etc.). The shuffled sequences were
always compared with the original one (without any shuffling). We note
that all our files have the same letter distribution. However only the
correlations on scales below the shuffling level are conserved. The
correlations (fluctuations) on higher levels which are based on the
large--scale structure of texts as well as on semantic relations are
conserved only in the original file. The objects of our investigation
were the book ``Moby Dick'' by Melville ($L\approx 1,170,000$ letters)
and the DNA--sequence of the lambda--virus ($L\approx 50,000$
letters).

\section{Entropy--like Measures of Sequence Structure}
Let $A_1 A_2 \dots A_n$ be the letters of a given substring of length
$n\le L$. Let further $p^{(n)}(A_1\dots A_n)$ be the probability to
find in a string a block with the letters $A_1\dots A_n$. Then we may
introduce the entropy per block of length $n$:
\begin{equation}
H_n=-\sum p^{(n)}(A_1 \dots A_n) \log p^{(n)}(A_1 \dots A_n)
\end{equation}
\begin{equation}
h_n = H_{n+1}-H_n
\end{equation}

Our methods for the analysis of the entropy of sequences were in
detail explained elsewhere~\cite{per}. We have shown that at least in
a reasonable approximation the scaling of the entropy against the word
length is for large n given by root laws of the type  

\begin{equation}
H_n = c_0 + c_1 \sqrt{n} + c_2 \cdot n ~.
\end{equation}
\begin{equation}
h_n = 0.5 \,c_1 \cdot n^{-\frac{1}{2}} + c_2~.
\end{equation}

In our earlier work on Moby Dick~\cite{epa} we assumed $c_2 = 0$ and obtained 
for our empirical data in the range $n = 10 - 26$ the constants $c_0 = 1.7$,
 $c_1 = 0.9$ . As shown already in 1951 by Shannon and a few years later by
Burton and Licklider, the limit of the uncertainty $h_n$ for large n which 
is given here by $c_2$ should be finite. With the estimate $c_2 = 0.05$ we 
have repeated our fit including data for Moby Dick obtained with a new 
method~\cite{per}. The parameters obtained by the new fit are
$c_0 = 1.7$, $c_1 = 0.5$, $c_2 = 0.05$. The dominating contribution ist given
by the root term, an observation first made by Hilberg. In this way the  
decay of uncertainties $h_n$ follows a scaling according to a power law.

We consider now a different approach to the entropy analysis which 
goes back to the work of Kolmogorov, Chaitin, Lempel and Ziv.
The algorithmic entropy according to Lempel and Ziv is introduced as
the relation of the length of the compressed sequence (with respect to
a Lempel--Ziv compression algorithm) to the original length.

\begin{figure}[htb]

  \centerline{\psfig{figure=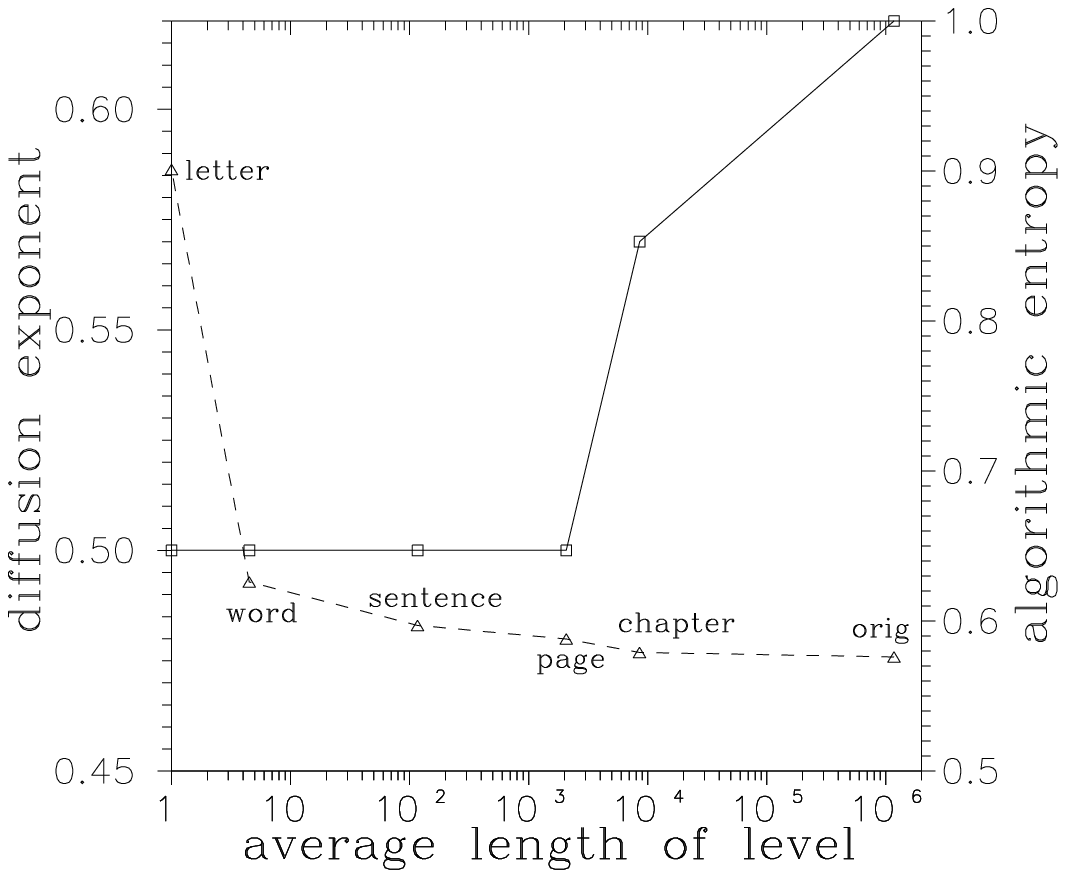,width=2.2in}\psfig{figure=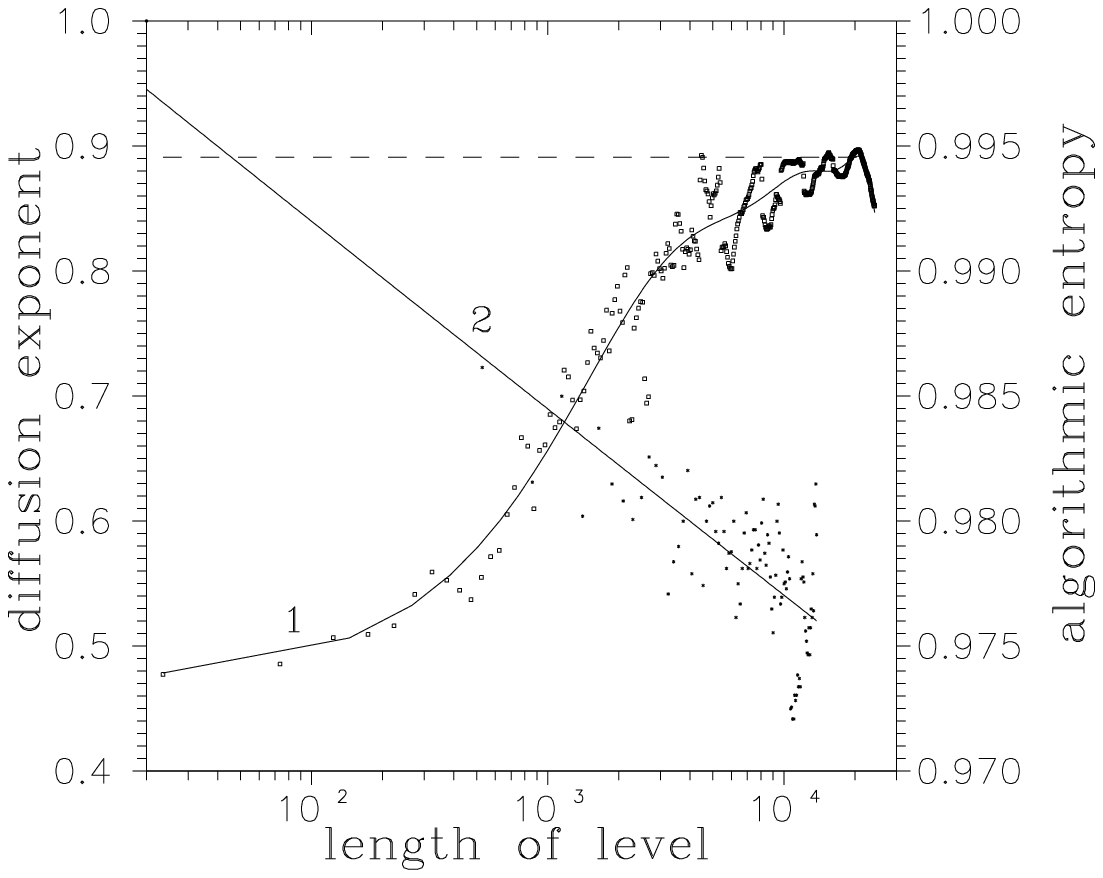,width=2.2in}}
\caption{Lempel--Ziv complexities (dashed line) and scaling exponents of 
  diffusion (full line) represented on the level of shuffling for the
  text Moby Dick (left) and for the Lambda--Virus DNA (right).}
\label{figmobycom}
\end{figure}

The results obtained for the Lempel--Ziv complexities (entropies) of
Moby Dick and of the lambda virus DNA are shown in Fig.~\ref{figmobycom}.
We see, that the Lempel--Ziv entropy of Moby Dick having a value of 
about 0.56 is much higher than than the Shannon entropies derived above.
This clearly shows that the compression algorithms based on the 
Lempel-Ziv method are by far not optimal. As show our shuffling 
experiments represented in Fig.~\ref{figmobycom}, compression algorithms
of Lempel-Ziv type are mainly based on rules below the page level.

\section{Correlation Functions, Power Spectra, Random Walk Exponents}
\begin{figure}[htb]
  \centerline{\psfig{figure=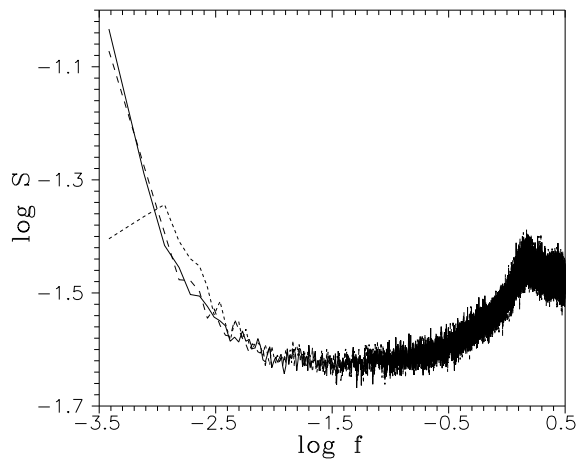,width=3.0in}}
\caption{Double--logarithmic plot of the power spectrum for Moby Dick. 
  The full line corresponds to the original text, the dashed line
  corresponds to the text shuffled on the chapter level and the dotted
  line to the text shuffled on the page level. Shuffling on a level
  below pages destroys the low frequency branch.}
\label{figmobyspec}
\end{figure}

In this part we closely follow the method proposed by Peng et
al.~\cite{peng,stan} and the invariant representation proposed
Voss~\cite{voss}. Instead of the original string consisting of
$\lambda$ different symbols we generate $\lambda$ strings on the
binary alphabet (0,1) ($\lambda = 32$ for texts). In the first string
we place a ``1'' on all positions where there is an ``a'' in the
original string and a ``0'' on all other positions. The same procedure
is carried also out for the remaining symbols. Then we generate random
processes corresponding to these strings moving one step upwards for
any ``1'' and remaining on the same level for any ``0''. The resulting
move over a distance $l$ is called $y(k,l)$ where $k$ denotes the
symbol. Then by defining a $\lambda$--dimensional vector space
considering $y(k,l)$ as the component $k$ of the state vector at the
(discrete) ``time'' $l$ we can map the text to a trajectory.  The
corresponding procedure is carried out for the DNA--sequences which
are mapped to a random walk on a $\lambda = 4$--dimensional discrete
space. The power spectrum is defined as the Fourier transform of the
correlation function $C(k,n)$ which measures the correlation of the
letters of type $k$ in a distance $n$~\cite{anishchenko}.  The
results of spectra calculations for the original file of the Bible,
for Moby Dick and for the same files shuffled on the word level or on
the letter level correspondingly were presented in a foregoing
work~\cite{ebelingneiman}. As a new result we present here the power
spectrum of Moby Dick shuffled on the chapter and on the page level
(Fig.~\ref{figmobyspec}).  \unitlength1.0cm

We see that the spectra of the original texts have a characteristic
shape with a well--expressed low frequency part. This shows the
existence of long--range correlations in Moby Dick. An estimate for
the exponent is 0.8 . However it is difficult to extract a precise
value for the exponent corresponding to a power law. One should better
look at the shape of the whole curve which reminds a piecewise linear
behaviour which is characteristic for multifractal structures. This
point certainly needs further investigations. However we believe, there
is no reason to expect that a long text as Moby Dick is a multifractal
in a quantitative sense. On the other hand the hierarchical character
of the structure of texts is beyond any doubt and we think that
spectral curves may be characterised as a quantitative measure of this
hierarchy.  Fig.~\ref{figmobyspec} shows that shuffling on the page
level already destroys the low--frequency branch of the spectrum. This
clearly proves that the origin of the $1/f$--fluctuations is on a
scale which exceeds the page level.  The contribution of high
frequencies corresponds to the structure on the word and sentence
level.  Let us study now the anomalous diffusion coefficients which
allow a higher accuracy of the analysis~\cite{stan}.  The mean square
displacement for symbol $k$ is determined as
\begin{equation}
\label{displ}
F^2(k,l)=<y^2(k,l)>-{(<y(k,l)>)}^2,
\end{equation}
where the brackets $<\cdot>$ mean the averaging over all initial
positions. The behaviour of $F(k,l)$ for $l\gg1$ is the focus of
interest. It is expected that $F(k,l)$ follows a power
law~\cite{stan}.
\begin{equation}
\label{diffusion}
F(k,l)\propto l^{\alpha(k)},
\end{equation}
where $\alpha(k)$ is the diffusion exponent for symbol $k$. We note
that the diffusion exponent is related to the exponent of the power
spectrum~\cite{stan}. The case $\alpha(k)=0.5$ corresponds to the
normal diffusion or to the absence of long--range correlations. If
$\alpha(k)>0.5$ we have an anomalous diffusion which reflects the
existence of long--range correlations.  Beside the individual
diffusion exponents for the letters we get also an averaged diffusion
exponent $\alpha$ for the state space.  The data are summarised in
Fig.~\ref{figmobycom}.

In the same way we can obtain other important statistical quantities:
higher order moments and cumulants of $y(k,l)$~\cite{ebelingneiman} By
calculations of the H\"older exponents $D_q$ up to $q=6$ we have
shown, that the higher order moments show (in the limits of accuracy)
the same scaling behaviour as the second moment. We repeated the
procedure described above for the shuffled files. A graphical
representation of the results for Moby Dick and DNA are given in
Figs.~\ref{figmobycom}.

We see from Figs \ref{figmobycom}--\ref{figmobyspec} 
that the original sequences show strong long--range correlations, i.e.
the coefficients of anomalous diffusion are clearly different from
$1/2$ and there exists $1/f$--noise.  After the shuffling below the
page level the sequences become practically Bernoullian in comparison
with the original ones since the diffusion coefficients decrease to a
value of about $1/2$ and there is no more $1/f$--noise. The decrease
occurs in the shuffling regime between the page level and the chapter
level. For DNA--sequences the characteristic level of shuffling where
the diffusion coefficient goes to 1/2 is about 500--1000. Our result
demonstrates that shuffling on the level of symbols, words, sentences
or pages, or segments of length 500--1000 in the DNA--case destroys
the long range correlations which are felt by the mean square
deviations.

\section{Conclusions}
Our results show that the dynamic entropies, the low frequency spectra
and the scaling of the mean square deviations, are appropriate
measures for the long--range correlations in symbolic sequences.
However, as demonstrated by shuffling experiments, different measures
operate on different length scales. The longest correlations found in
our analysis comprise a few hundreds or thousands of letters and may
be understood as long--wave fluctuations of the composition. These
correlations (fluctuations) give rise to the anomalous diffusion and
to long--range $1 / f$--fluctuations. These fluctuations comprise
several hundreds or thousands of letters. There is some evidence that
these correlations are based on the hierarchical organisation of the
sequences and on the structural relations between the levels. In other
words these correlations are connected with the grouping of the
sentences into hierarchical structures as the paragraphs, the pages,
the chapters etc. Usually inside certain substructure the text shows a
greater uniformity on the letter level. In order to demonstrate this
we have shown in earlier work~\cite{ebelingneiman} the (averaged over
windows of length 4000) local frequency of the blanks (and other
letters) in the text Moby Dick in dependence on the position along the
text. The original text shows a large--scale structure extending over
many windows. This reflects the fact that in some part of the texts we
have many short words, e.g. in conversations (yielding the peaks of
the space frequency), and in others we have more long words, e.g. in
descriptions and in philosophical considerations (yielding the minima
of the space frequency). The shuffled text shows a much weaker
non--uniformity of the text, the lower the shuffling level, the larger
is the uniformity. More uniformity means less fluctuations and more
similarity to a Bernoulli sequence. For the case of DNA--sequences no
analogies of pages, chapters etc. are known. Nevertheless the reaction
on shuffling is similar to those of texts. Possibly a more careful
comparison of the correlations in texts and in DNA sequences may
contribute to a better understanding of the informational structure of
DNA.

\end{document}